# The Social Contract for AI

Mirka Snyder Caron[1], Abhishek Gupta[2]
[1]Associate, Montreal AI Ethics Institute
[2]Founder, Montreal AI Ethics Institute, Machine Learning Engineer, Microsoft
mirka@montrealethics.ai , abhishek@montrealethics.ai

## Abstract

Like any technology, AI systems come with inherent risks and potential benefits. It comes with potential disruption of established norms and methods of work, societal impacts and externalities. One may think of the adoption of technology as a form of social contract, which may evolve or fluctuate in time, scale, and impact. It is important to keep in mind that for AI, meeting the expectations of this social contract is critical, because recklessly driving the adoption and implementation of unsafe, irresponsible, or unethical AI systems may trigger serious backlash against industry and academia involved which could take decades to resolve, if not actually seriously harm society. For the purpose of this paper, we consider that a social contract arises when there is sufficient consensus within society to adopt and implement this new technology. As such, to enable a social contract to arise for the adoption and implementation of AI, developing:
1) A socially accepted purpose, through
2) A safe and responsible method, with
3) A socially aware level of risk involved, for
4) A socially beneficial outcome, is key.

## 1 Introduction

Like any technology, AI systems come with inherent risks and potential benefits. It comes with potential disruption of established norms and methods of work, societal impacts and externalities which at times can be foreseen, and at times, cannot. Technology has often replaced or changed the human labor or other technologies considered obsolete when compared to the alternative, more efficient, more convenient one.

One may think of the adoption of technology as a form of social contract, which may evolve or fluctuate in time, scale, and impact. It is important to keep in mind that for AI, meeting the expectations of this social contract is critical, because recklessly driving the adoption and implementation of unsafe, irresponsible, or unethical AI systems may trigger serious backlash against industry and academia involved which could take decades to resolve, if not actually seriously harm society.

For the purpose of this paper, we consider that a social contract arises when there is sufficient consensus within society to adopt and implement this new technology. And this adoption can at times take many years, if not decades, to become the most common method of work, and at times it can trigger a snowball effect and have potential global adoption and impact in a matter of days.

More often than not, it is society that decides whether to accept the effects and societal impacts of a technology, and consequently adopt, implement or use this technology, because it is found useful, safe, and convenient, (and perhaps entertaining). It is often optimal to let society develop its own voluntary curiosity and awareness, to learn about it, understand it, try it out and test it, and finally accept to use it because it has been found to bring positive effects for the individual, the family, the business, the community, and for the general collective.

An issue arises when such adoption is imposed on society, for instance through "stealth mode" application with poor transparency, disclosure, or insufficient public education and awareness. This may trigger social and political instability, as well as difficulties in consumer market access and penetration, when it is perceived that freedom, self-determination, human rights and fundamental values are being hard pressed by political or corporate agendas which are not sufficiently aligned with public feeling, public feedback, or individual and society's interests.

Technological change, even change in general, can be difficult. If it is perceived that the new technology is complicated to understand or to use, that its real-felt benefits or positive effects are not clearly identified or hard to conceptualize, that the technology can be difficult to control or can provide for unforeseeable or unpredictable scenarios, or if there is fear of perceived or actual risk that harm may instantly occur from this technology and that voiced concerns are unheard by official representatives, or society members sense they are powerless in influencing the course of devel-opments or regulation, adoption and implementation may inevitably lag.

As such, to enable a social contract to arise for the adoption and implementation of AI, developing:

1) A socially accepted purpose, through
2) a safe and responsible method, with
3) a socially aware level of risk involved, for
4) a socially beneficial outcome, is key.

## 2 A Socially Accepted Purpose

Clear identification of the purpose of an AI system ought happen at time of design and prior to deployment and scaling. Identification ought be done both technically and through governance policy, in explicit, unambiguous and clear human language. We do not mean a socially accepted purpose to be exclusively intended for non-profit application or social welfare measure. At the "hard" core of a socially accepted purpose, it is "common sense" that such purpose would, at a minimum, meet existing human rights and constitutional and fundamental and ethical values of a society.

On a larger scale, compliance to other existing regulation and regulatory guidelines and directives in given industries is also obvious, as regulation intends to frame acceptable behaviors from unacceptable ones, and as it ought have been harmonized, verified against, and been drafted in compliance with human rights and fundamental values charters.

It is important to keep in mind that it would be inappropriate to consider a society as a conglomerate of organized individuals with fixed, unchanging values and interests. These evolve and fluctuate in time and in geography. Varying levels of trust and optimism in AI across different societies ought be accounted for and respected, despite international pressure and politicized national investment strategies. As such, bridging the gap between public feeling and national policy by obtaining public feedback and voluntary participation and cooperative dialogue would be most optimal to ensure purposes remained aligned with evolving social values.

In the case of online-learning systems, there is also a potential for drift in the output probability distributions over time as the system interacts with real-world data such that the system veers away from the identified, socially-accepted purpose and there needs to be instrumentation to monitor, alert and act on such deviations.

## 3 A Safe and Responsible Method

As human rights and values charters, and regulatory frameworks often already exist, although some could be improved to better align with impacts of ongoing technological developments, the question remains as to the manner in which to increase the efficient combination of regulatory, social, corporate and technical measures to ensure purpose of the AI remains safe and responsible.

The goal, output and resulting outcomes of an AI system can be technically embedded, to a certain extent, as to align with such human rights, values, and safety considerations, although research is still ongoing pertaining to embedding morality within algorithms and as such, other complimentary governance measures must accompany the existing technical security measures to ensure purpose remains socially accepted.

Some of the other potential areas that can help in ensuring safe and responsible methods, or at least provide a framework within which such methods can be evaluated are: technical standards and certifications, regulatory templates for data licensing, regulatory guidelines and legislative amendments from governing bodies both at an industry and government level, local human rights based due diligence (local to encourage contextual and cultural sensitivity to keep in alignment with social acceptance which might be harder with internationally mandated human rights frameworks). Interruptibility also offers a pathway to providing more human-in-the-loop style controls in cases of the AI-enabled system going out of their designated, socially accepted purposes.

## 4 A Socially Aware Level of Risk involved

As said prior, any technology has an inherent level of safety risk for society. Depending on purpose and method of application, this risk runs across an acceptability spectrum. In other words, for technology to exist, it is almost impossible to implement a zero-risk policy without completely stifling innovation in this technology.

Consequently, it is more a question of how much risk a society is willing to accept for the potential benefits of a given technology. At times, national and international agreements will severely restrict or limit the development of specific technologies as they are deemed high-risk, for instance weapons falling under "Mutually Assured Destruction" qualifications.

In other words, safe AI does not mean the need to guarantee a continuous zero-risk AI. However, there are generally very high social expectations that a product will be "fit for purpose", with low or very low probability of danger, hazard or harm, when it is purchased or used. For products which may be harmful for human use or consumption, adequate warning and disclosure of such risks, as well as instructions for use and for non-use scenarios, ought be brought to the knowledge of the user. If an AI system must be used in a specific context only, then this also ought be clearly indicated, as to ensure that usage is governed according to developers purpose.

As every member of a society would not have the expertise to deeply understand the complexities of the underlying models, factors, variables, diverse classes of algorithms used and their limitations, there are inherent social expectations that AI systems ought to be made available for use when it is safe and beneficial to use. It is essential in such a case to also make public the context and intended purpose of use of the AI system and the parameters under which this system was tested and developed to level-set how users interact and utilize the system.

As such, there is often a social intent to re-allocate at least 1) partial accountability and responsibility on designers, programmers and product manufacturers with the appropriate competence and expertise to develop safe AI systems, as well as 2) supervision, monitoring, and enforcement to designated regulators, appropriately supported by 3) punitive and compensatory legal mechanisms available under legal recourse for damages or other remedial and punitive measures through the public justice system, in application of possible tort law, contractual law, consumer protection, or regulatory non-compliance indemnification. Compensation may also be made available through private settlement processes such as an independent ombudsman, or through compensation or insurance schemes.

## 5   A Socially Beneficial Outcome

Ultimately, putting together the socially accepted purpose through a safe and responsible method while being socially aware of the risk involved, the AI-enabled system must produce a socially beneficial outcome that is in line with the context and culture of the target audience. When thinking about using AI-enabled systems in a social good context or in general thinking about AI-enabled systems that are used in society in a safe and responsible manner, it can often be hard to define what a socially beneficial outcome could be. Given varying self-interests of groups within society, unless the system is developed for a very small target niche there will be many trade-offs that need to be made when such a system is deployed for many, possibly heterogeneous groups.

Yet, these are not hitherto unseen problems - for any large-scale public systems that are deployed in society, they affect people across many different factions and such discussions need to be had in those cases as well on the trade-offs to achieve a socially beneficial outcome. Decisions are made in that case on the "who" for those outcomes and the definitions of "socially beneficial" as it applies to that target audience.

We believe that just as in software development there is user research and feedback in the prototype stages to determine if the product development is on track to meet the needs of potential customers, there is a need and urgency to have public discussions on the development and deployment of AI-enabled systems like facial recognition to transparently determine the "who" that benefits from it and what the "socially beneficial" outcome is and what the unintended consequences might be in the use of such a system. Public consultations serve as a great mechanism in eliciting an understanding of what is relevant from a contextual and cultural standpoint for that community and at the same time it empowers the very people that are going to be subjects of these systems in having a voice in the technical and policy measures that surround the development and deployment of these pieces of technology.

## 6   Considerations

Please note that we include in the present definition of artificial intelligence ("AI") automated decision-making systems, machine-learning and predictive algorithms, whether rendered physically autonomous in the form of robots or not. Also, please note that when we use "socially" or "society", we do not fall into the semantics of what constitutes society or not, and what types of societies there are, but we are aware and pertain that the individual and collective interests need be weighted, although order of priority of such weights may vary on a spectrum based on different psycho-social and geographical factors. Finally, although much could be said about military applications of AI, the scope of this paper excludes these particular considerations.

## 7   Conclusion

In this paper, we propose a high-level, flexible framework that can be utilized throughout the development and deployment lifecycle of an AI-enabled system to ensure that it indeed is safe, inclusive and ethical while at the same time producing social good as a consequence of its utilization. We highlight some key existing techniques and measures that can be adopted to implement the framework while also indicating some potential future areas of work that can add more detail and concreteness to the implementation. It serves as a viable starting point for organizations that are developing these systems to meet their responsibilities in developing safe, inclusive and ethical AI systems while orienting their systems towards producing socially beneficial outcomes. Simultaneously, it provides the users, other public organizations including regulatory bodies, NGOs, governments, etc. the same high-level framework to evaluate the systems. The diversity in underlying mechanisms to implement the framework can serve as a counterbalance system to analyze these systems from different angles ultimately helping to generate more robust and meaningful analyses of the systems.